# Trending Videos: Measurement and Analysis

Iman Barjasteh
Michigan State University
428 S. Shaw Lane
East Lansing, MI 48824
barjaste@msu.edu

Ying Liu
Michigan State University
428 S. Shaw Lane
East Lansing, MI 48824
liuying5@egr.msu.edu

Hayder Radha
Michigan State University
428 S. Shaw Lane
East Lansing, MI 48824
radha@egr.msu.edu

## ABSTRACT

Unlike popular videos, which would have already achieved high viewership numbers by the time they are declared popular, YouTube trending videos represent content that targets viewers' attention over a relatively short time, and has the potential of becoming popular. Despite their importance and visibility, YouTube trending videos have not been studied or analyzed thoroughly. In this paper, we present our findings for measuring, analyzing, and comparing key aspects of YouTube trending videos. Our study is based on collecting and monitoring high-resolution time-series of the viewership and related statistics of more than 8,000 YouTube videos over an aggregate period of nine months. Since trending videos are declared as such just several hours after they are uploaded, we are able to analyze trending videos' time-series across critical and sufficiently-long durations of their lifecycle. In addition, we analyze the profile of users who upload trending videos, to potentially identify the role that these users' profile plays in getting their uploaded videos trending. Furthermore, we conduct a *directional-relationship* analysis among all pairs of trending videos' time-series that we have monitored. We employ Granger Causality (GC) with significance testing to conduct this analysis. Unlike traditional correlation measures, our directional-relationship analysis provides a deeper insight onto the viewership pattern of different categories of trending videos. Our findings include the following. Trending videos and their channels have clear distinct statistical attributes when compared to other YouTube content that has not been labeled as trending. Based on the GC measure, the viewership of nearly all trending videos has some level of directional-relationship with other trending videos in our dataset. Our results also reveal a highly asymmetric directional-relationship among different categories of trending videos. Our directionality analysis also shows a clear pattern of viewership toward popular categories, whereas some categories tend to be isolated with little evidence of transitions among them.

## 1. INTRODUCTION

YouTube as a user generated content is one of the largest and most popular video sharing websites. It hosts over four billion views a day. YouTube provides public statistics regarding its uploaded videos, most notably the number of views, which shows the aggregate number of times a video has been watched up to that point. Naturally, the number of views for a video indicates the level of popularity of that video; and it takes a varying amount of time for a video to become popular (if it becomes popular). Meanwhile, there are some videos that can attract users' attention in a relatively short time. YouTube also supports a feature called trending, which represents content that has the *potential* of becoming popular in a relatively short time. Consequently, although trending videos are usually not popular (yet) when declared as trending by YouTube, they have the potential of becoming popular (eventually). For example, some videos are labeled trending while having only few hundreds in viewership numbers. From another perspective, through trending videos, YouTube tries to highlight emerging trends developing within different viewership communities.

Meanwhile, the general attributes of the viewership of trending videos have not been studied thoroughly. To the best of our knowledge, basic statistics about YouTube trending videos have not been studied, analyzed, or even received any adequate attention. Considering the fact that more than one billion unique users visit YouTube each month and they upload 72 hours of video every minute [26], YouTube is the best place for e.g. brand engagement or advertising, but it is genuinely difficult and competitive to get the attention of users. Therefore when a video becomes popular, it is exposed to millions of users for free and has the opportunity of keeping their attention for a while. Finding these trends are significantly important that many different websites have been emerged just to pick up on the latest trends on the web, such as "buzzFeed" or "whatTheTrend". There are other kinds of websites such as "channelFactory" that tries to build an audience on YouTube for content owners or advertisers. Better understanding of YouTube trending videos and their statistics, and a deeper insight about their lifecycles, can greatly affect the strategies for marketing, target advertising, recommendation systems and search engines, as was suggested by prior YouTube measurement studies [2]. This represents a key motivation for our effort.

In this paper, we present our findings for measuring, analyzing, and comparing key aspects of YouTube trending videos. Our study is based on collecting and monitoring high-resolution time-series of the viewership and related statistics of more than 8,000 YouTube videos over an aggregate period of nine months. To put this number into perspective, YouTube declares only tens of videos as trending on a daily basis. This number is highly selective when compared to the thousands of videos uploaded on YouTube every minute [26]. Furthermore, and from the onset of our study, our goal has been to collect high-resolution viewership time-series of trending videos to achieve robust analysis of their entire lifecycles. Since trending videos are declared as such just several hours after they are uploaded, we are able to analyze trending videos across their lifecycle; this provides an invaluable insight into their viewership time-series over a critical period of their lifetime. The main contributions of this paper can be summarized as follows:

1. *Analysis of the viewership lifecycle and basic statistics of trending-video content*. First, we analyzed and evaluated a variety of statistics of a comprehensive dataset of about 4,000 trending videos that we monitored over a relatively long time (more than two months) using the YouTube data application programming interface (API). This dataset represents traditional statistics regarding the content itself; this includes number of views, number of comments, time durations of each video clip, categories of these video clips, and related viewership statistics. Our initial objective was to answer basic questions about trending videos: What does a lifecycle of a trending-video viewership look like? How long does it take for a trending video to become popular (if it becomes popular)? What is the percentage of trending videos that do become popular? What are the categories (e.g., entertainment, education, news, etc.) and clip durations of trending videos?

2. *A comparative analysis of trending and non-trending videos.* Second, we aimed at identifying any salient differences in the statistical properties between trending videos and other videos that are not labeled trending by YouTube (we refer to the latter type as "non-trending" videos; although they may become popular). To that end, we needed to collect data for non-trending videos that we could monitor over the same time span while monitoring a corresponding set of trending videos. Consequently, and for the sake of identifying distinguishable statistical attributes of trending videos over time, we monitored and collected all relevant statistics of more than 2,000 *recently uploaded* YouTube videos and compare them with the same number of (newly collected) trending videos that we monitored over exactly the same time period. Hence, we have a second dataset of more than 4,000 YouTube videos that consists of more than 2,000 of recently-uploaded (non-trending) videos and a similar number for trending videos. Similar to trending videos, the recently-uploaded content provided us with the opportunity of collecting and monitoring their statistics from the time they were uploaded.

3. *Analysis of the profile of trending video uploaders*. In addition to monitoring viewership statistics about the content itself (as explained above), we also had the objective of gaining insight at other factors that might influence the reason why a trending video is labeled as such. These factors might not be very obvious to (or arguably hidden from) a casual viewer. In particular, we collected statistics about the users who upload trending videos (i.e. the content providers); this includes the view count and subscriber count of uploaders' channels, and other uploaders' statistics. We conducted a comparative analysis between trending video uploaders' profile and the profile of uploaders of *recently uploaded* video. We believe that this comparative analysis sheds some light on some of the factors that might be influencing the determination and popularity of trending videos. By measuring, analyzing, and comparing key statistics regarding trending videos and their uploaders, we believe that this part of our work provides an insight into some of the properties of this crucial class of YouTube content.

4. *Directional-Relationship analysis of trending-video viewership and the viewership of its categories*. Another objective of our work is to study and analyze the *directional-relationship* among the viewership time-series of trending videos, in general, and of those trending videos that become popular in particular. Unlike traditional correlation analysis, which only quantifies the strength of relationship between two processes, our directional-relationship analysis reveals both the direction and strength of interactions between any pair of random processes. We have employed Granger Causality (GC) with significance testing to measure such directional-relationship, not only between each pair of viewership time-series, but more importantly among the viewership of the different categories (e.g., entertainment, sports, news, music, etc.) of content offered by YouTube trending videos. Hence, this form of directional-relationship analysis can provide an insight onto the viewership pattern among these different categories.

We believe that our study represents an important (arguably first) attempt for developing an insight into trending videos and the viewership pattern among its different categories. Consequently, we hope that the insight provided by our study will inspire future research into trending videos, and it might impact key applications such as marketing and advertisement. For example, YouTube has more than a million advertisers that employ Google ad platforms, and according to YouTube stats [26] the majority of these advertisers are small businesses. Having an insight about trending videos' lifecycles and the viewership pattern among its popular categories can arguably be instrumental for such application.

The remainder of the paper is organized as follows. In section II, we discuss prior related work. In section III, we describe our data collection effort. In section IV, we present and analyze different statistics related to trending videos, including a comparative analysis with non-trending videos, and analysis of the uploaders' profile. In section V, we present our directional-relationship analysis. Section VI concludes the paper.

## 2. RELATED WORK

There have been several studies conducted on YouTube due to the fact that it is (one of) the most popular video sharing website(s). These studies have focused on different characteristics of videos and user profiles. In [1], Zhou et al. studied the impact of YouTube recommendation system on video views. In [2], Cha et al. analyzed the popularity life-cycle of videos, the intrinsic statistical properties of requests and their relationship with video age. In [6], Davidson et al. studied the video recommendation system that YouTube uses and its role in increasing the total number of views for videos. In [7], Cheng et al. studied the statistics and social network of YouTube videos. They found that the links to related videos generated by uploaders' choices have clear small-world characteristics. Ding et al. studied YouTube uploaders and demonstrated the positive reinforcement between on-line social behavior and uploading behavior in [4]. Pattern of influence in individual recommendation network has been studied by Leskovek et al in [8].

Several previous works studied the impact of YouTube recommendation system and uploaders on total view count of videos. There are some other works focusing on the impact of videos' categories or the size of YouTube. For example in [3], Filippova et al. studied the video categories and considered the task of assigning categories to YouTube videos based on the text information related to videos, such as video title, description, user tags and viewers' comments. In [5], Zhou et al. studied the counts of YouTube videos via random prefix sampling. They designed an unbiased estimator of the total number of YouTube videos.

Recently, some work has been pursued to capture trends in social media. Reed developed a standard system for detecting emerging trends in YouTube video posts [15]. Asur et al. provided a theoretical basis for analyzing the formation, persistence and decay of trends for the trending topic on Twitter [9]. However, to the best of our knowledge, YouTube trending videos have not been studied thoroughly.

The diffusion or dissemination model of video onto the market has also been studied [22][23][24]. Broderson et al. investigates the relationship between popularity and locality of YouTube videos and analyze how the geographic properties of a video's views change over its lifetime [25].

## 3. DATA COLLECTION

YouTube provides a data API that allows a systematic collection of public standard feeds and statistics related to videos and users. We collected our datasets through this YouTube data API. We focused on collecting data on trending videos, which is a recently added feature to the API. For our comparative analysis, we also collected data regarding other (non-trending) videos using another feed made available by YouTube for recently-uploaded content as we explain further below.

The API places some restrictions on data collection. In particular, one cannot record an accurate statistics for the viewership count as a function of time over the past history of a desired video by simply accessing the aggregate viewership statistic. To gain an accurate viewership count over time, and hence, to generate a valid time-series of viewership statistics one needs to monitor the desired video (virtually continuously) over time. Otherwise, there is no way of acquiring any accurate history for the number of times a video has been viewed.

The YouTube API supports a feed called *on_the_web*, which lists and provides access to trending video statistics. To collect data about these trending videos using on_the_web feed, we could only retrieve about 200 new trending videos everyday by sending a request to the API; and after removing duplicate videos, which were initially collected by our software, we usually had more than 100 (but less than 150) new and unique trending videos every day; and for each batch of new and unique trending videos we collected every day we added them to our existing video pool. Once we identified a new trending video, we began collecting basic statistics such as its number of views and number of comments. YouTube feeds are updated periodically with different update frequencies for different kind of feeds. YouTube feeds such as most viewed videos or top favorite videos are updated every 30 to 40 minutes [17]. Statistics for a video, such as number of views, number of available comments, or its rating are updated anywhere from 30 minutes to two hours. However, depending on the server load or for videos that are viewed infrequently, updates occur less frequently [17]. In our case, and to ensure that we keep track of our feeds with high resolution, especially the number of views, we collected our data with a frequency of every 40 minutes for each video.

The first part of our study is based on monitoring the viewership and related statistics of about 4,000 trending videos (3922 to be exact) over two months between May and July of 2012. For this initial 4,000 video dataset, and for each video in this dataset, we had a minimum of two-week history and up to two-month data of its viewership statistics. This was crucial for our initial assessment of some of the basic analysis and characteristics of trending videos, and especially regarding the development of a clear understanding of the nature of their lifecycle (e.g., how popular trending videos become, how long does it take them to reach a certain popularity level, etc.). As we show in this paper, trending videos that become popular, tend to reach significant popularity levels within a week or two; afterword, the rate at which their viewership increases tends to saturate, as the case with many popular videos.

The YouTube API also supports a feed called *most_recent*, which provides access to statistics regarding videos that are uploaded recently. Collecting data about *recently uploaded* videos through the most_recent feed is quite similar to collecting trending videos through the on_the_web feed. The daily rate of acquiring new and unique recently-uploaded videos through the most_recent feed was more than 100 videos (i.e., very similar to the number of unique and new trending videos that we were able to identify on a daily basis). It is important to note that the recently-uploaded videos we were monitoring through the most_recent feed were different from the videos that we were monitoring through the on_the_web feed. Hence, in this paper, we sometime refer to the recently-uploaded videos that we monitored as *non-trending* videos mainly to distinguish them from trending videos. Our objective for collecting data about non-trending videos is to provide a comparative analysis about some of the key statistics of both types of content and their uploaders. We needed to collect these two types of time-series over the same time duration. Consequently, and in addition to the initial set of dataset of 4,000 trending videos that we mentioned above, we collected a second dataset containing trending and recently uploaded videos. We simultaneously collected this newer data over four weeks to create an impartial dataset containing 2,000 videos for each feed (on_the_web and most_recent). Therefore, the total number of videos we collected and monitored is around 8,000. We should note that we collected data for all trending videos that were made available by the YouTube API. From that perspective, the collected data represents the ground truth of YouTube trending videos.

Furthermore, we retrieved data related to the user who uploaded each video. For each uploader, we retrieved some feeds, such as gender, the view count of user's channel, the subscriber count of user's channel, and the total upload views. Further details about our datasets are provided in the different analysis sections.

## 4. LIFECYCLE ANALYSIS

In this section, we try to shed some light into two questions about trending videos: How popular trending videos become? And how long does it take them to reach a particular popularity level? Answering these questions provide an insight about the characteristics of the lifecycle of trending videos.

Before proceeding, it is crucial to highlight the time-dependency of our dataset of 4,000 trending videos. As we mentioned in the Data Collection section, we collected our data over a two-month period starting from May 2012; and each day we were adding a new collection of trending videos to our dataset. For the purpose of analysis, we declared a certain point-in-time in our data-collection process as an end-time (mid July 2012), mainly as a reference point. Based on this reference end-time, the amount of time over which we can conduct our analysis of these videos depends on the time when they were collected. For example, the video we began monitoring during the first two weeks of our study (in May 2012), can be analyzed over a minimum of 60 days; and videos we began monitoring during the second two-weeks of our study, can be analyzed over a minimum of 45 days; and so on. Consequently, we divide our dataset of 4,000 videos (3922 to be exact) into four (progressively inclusive) groups. The largest subset, which includes all 3,922 videos, can be analyzed over a minimum of 15 days; the second subset represents 3,330 videos that can be analyzed over a minimum of 30 days; and so on as shown in Table 1. Note that the subset of 1,115 videos we collected during the first two-weeks can be analyzed over the longest period of 60 days. These 1,115 videos are included in the larger inclusive set of 2,232, which can be analyzed over a minimum of 45 days. We refer to these four groups as Inclusive sets I1, I2, I3, and I4 (see Table 1). Here, I1 $\subset$ I2 $\subset$ I3 $\subset$ I4. From these inclusive sets, we can define corresponding Exclusive sets E1, E2, E3, and E4. For example, E3 is the set of videos that we collected during the third two-week period of our study. These videos can be analyzed over 30 days similar to the inclusive set I3. (Note that each exclusive set is a subset of its corresponding inclusive set; for example, E3 $\subset$ I3.) While I3 $\subset$ I4, the exclusive set E3 does not belong to any other exclusive sets, and it does include any videos that belong to the other exclusive sets. Note that the first inclusive set (I1) is the same as the first exclusive set (E1).

**Table 1: The number of videos in each analysis subset**

| **Minimum days over which can be analyzed** | 60 days | 45 days | 30 days | 15 days |
|---|---|---|---|---|
| **Number of videos (Inclusive set#)** | 1,115 **(I1)** | 2,232 **(I2)** | 3,330 **(I3)** | 3,922 **(I4)** |
| **Number of videos (Exclusive set#)** | 1,115 **(E1)** | 1,117 **(E2)** | 1,098 **(E3)** | 592 **(E4)** |

Now, we address the first question: How popular trending videos become? We evaluated the distribution of the popularity of our all trending videos in our dataset (i.e., I4 in Table 1) at the end of our analysis time (mid July 2012).

The histogram for this distribution is shown in Fig. 1(a). The figure shows that about 8% of videos (321 of them) in our dataset achieved more than one million views; and more than 40% have achieved a minimum of 100,000. It is clear that trending videos receive a wide range of popularity levels. We also recorded the same distribution of our dataset four months later (mid November 2012) and eight months later (mid March 2013) and the results are shown in Fig. 1(b) and 1(c) respectively. Although one can see some increase in the percentage of more popular videos, it is clear that there are very small changes in the overall shape in the viewership distribution over these periods. These observations have an impact on other aspects of our analysis.

One of the potential issues with the viewership distribution shown in Fig. 1 is that it includes videos with different lifetimes (from 15 days to more than 60 days due to the timeline nature of the comprehensive set I4). To gain a better insight into the impact of the lifetime of videos on their viewership, we evaluated the cumulative viewership for the four different exclusive sets (E1 to E4 in Table 1).

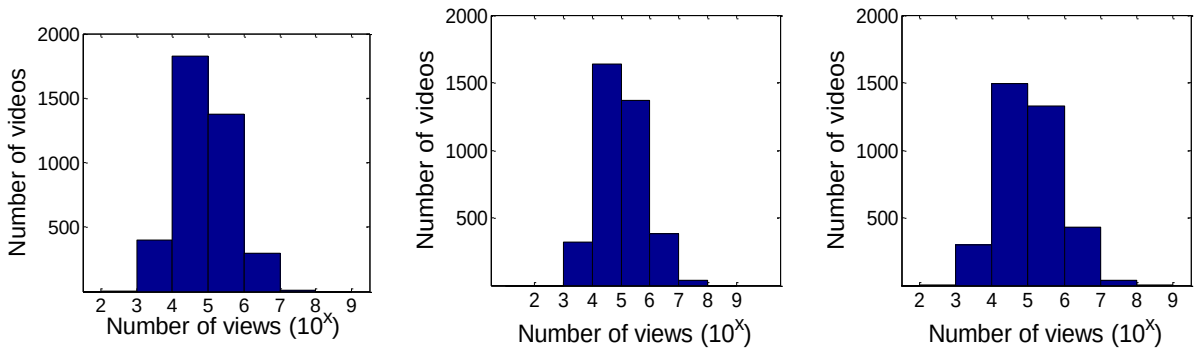

**(a)** **(b)** **(c)**

**Figure 1. Distribution of the final aggregated viewership of all 3922 videos in our dataset of Table 1 as recorded (a) on July 11, 2012 and (b) on November 14, 2012 and (c) on March 21, 2013.**

The results are shown in Fig. 2. For each day in this figure, we averaged the cumulative viewership of all videos (within the corresponding set) based on their age relative to the first day of their lifetime. Despite some variation among the distributions of the four sets, these plots illustrate a consistent pattern of steep increase in viewership videos at the early stages of their lifetime; then, there is a tendency of slowing down in terms of the rate of increase in cumulative viewership.

To further examine the rate of change in viewership over the lifetime of trending videos, we divided each exclusive set (E1 to E4) into different subsets depending on their final popularity level (i.e. according to their popularity at the end-time of our analysis). Here, we divide each set into five popularity levels starting from the range $10^3$-$10^4$ views and ending with the range $10^7$-$10^8$. Fig. 3 shows the cumulative viewership for the four sets (E1 to E4) over the durations of time over which they could be analyzed.

Fig. 3 reinforces the observation that trending video's viewership cycle tends to slow down rather quickly after the early stages of their lifetime; this observation seems to be true regardless of the final popularity level achieved by these videos. Overall, we observed a slowdown in viewership takes place about two weeks after a trending video is uploaded, virtually independent of its final popularity level. To illustrate this point further, Fig. 4 shows the times ($x$-axis) when a certain percentage of final viewership ($y$-axis) is reached for the inclusive sets (I1 to I4). For example, the plot associated with I4 (the whole 3,922 videos in our dataset), shows the percentage of viewership achieved over the first 15 days of the lifetime of all videos in that set. It is clear from this figure that these trending videos achieved about 80% of their popularity within the first two weeks of their lifetime.

Comparing the complete lifecycles of trending videos with that of non-trending videos, which "eventually" become popular is interesting, but it is virtually impossible since by the time non-trending videos become popular accurate data will not be available anymore. To address this issue, in the following section we compare trending videos with recently-uploaded videos, which enable us to monitor their viewership and other statistics from the early stages of their lifecycle in a similar manner to trending videos.

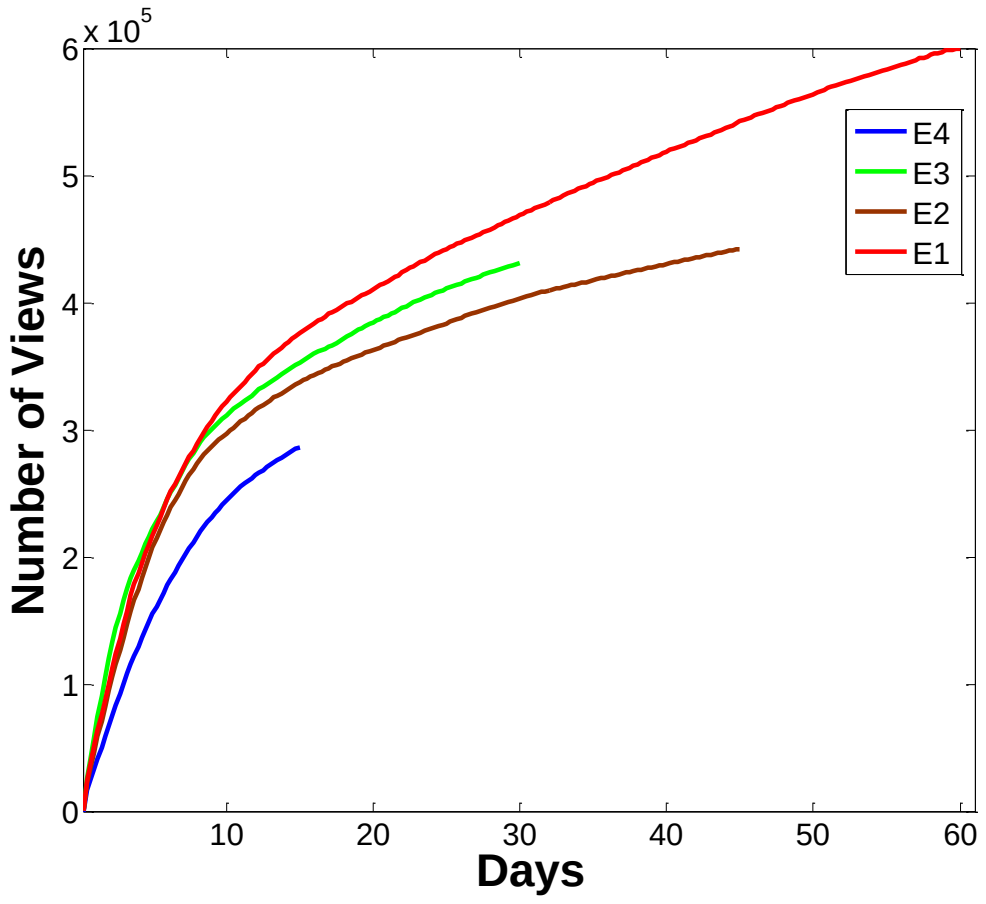

**Figure 2. Mean of accumulative viewership time-series of trending videos in four subsets.**

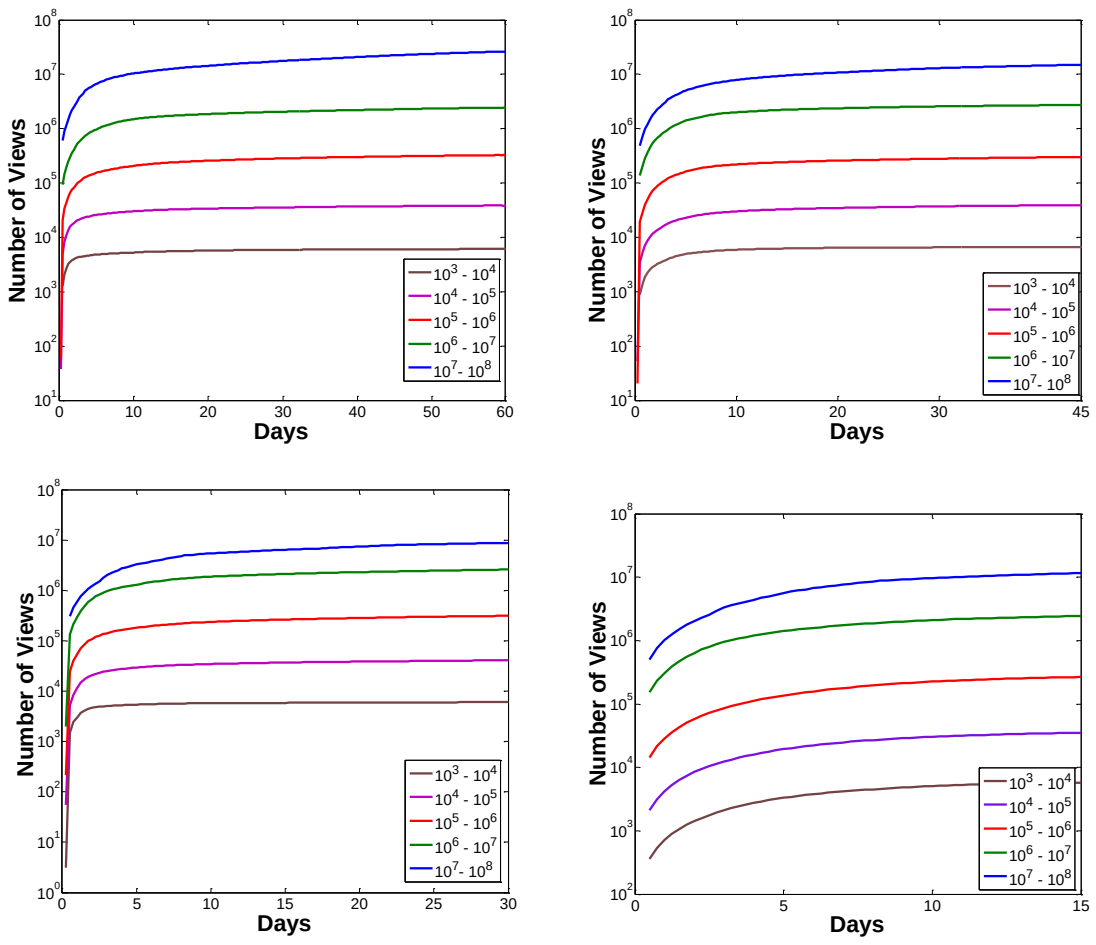

**Figure 3. Cumulative viewership of the four exclusive sets (E1 to E4 from top-left to bottom-right) over their corresponding lifetime in our dataset. Each set divided into five popularity levels depending on the final viewership numbers achieved at the end of our analysis time.**

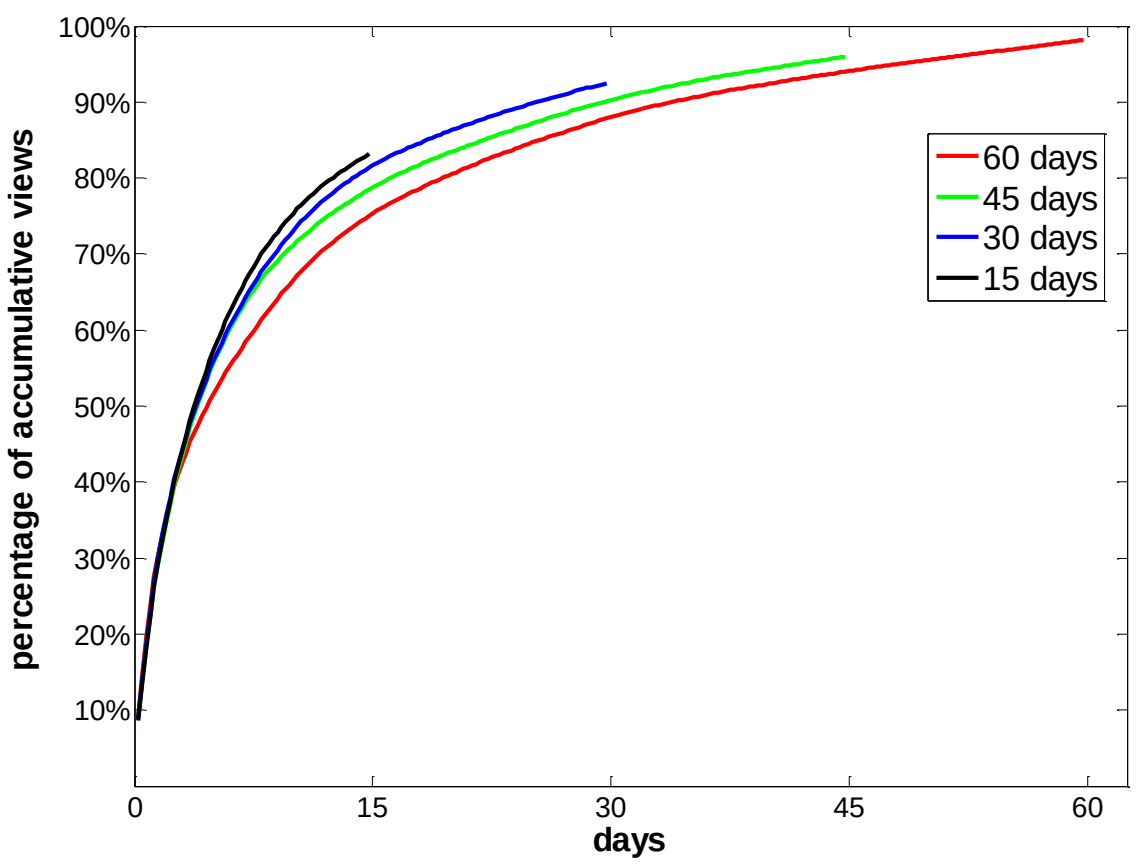

**Figure 4. Percentage of accumulative viewership over time**

# 5. TRENDING VERSUS NON-TRENDING

In this section, we compare some of the attributes of trending and non-trending videos with the objective of identifying any potential distinct differences between them. As described in the Data Collection section, for this part of our effort, we collected data for more than 2,000 recently-uploaded videos and about the same number of trending videos, all of which we monitored over the exact same duration of around four weeks. We conducted two types of comparative analyses. The first one is regarding standard video feeds, which provide data about basic statistics such as viewership numbers, duration-of-time of each clip, number of comments a video receives, etc. The second analysis is regarding the profiles of the videos' uploaders, such as their gender, their standing with YouTube, and their channels' characteristics. The following two subsections present our findings for these two analyses.

## 5.1 Comparative Analysis of Video Feeds

### *5.1.1 Durations*

Our dataset of trending videos and recently uploaded videos consist of 16 different categories by default as defined by YouTube. The time-duration mean for each category of videos are shown in Fig. 5 for both trending and recently-uploaded videos. We observe that the mean of the videos' length vary significantly according to their categories. There is also a clear indication that trending videos tend to have longer duration under virtually all categories (except for one). The distribution of all videos based on their durations for both datasets are shown in Fig. 6(a). We observe that around 30% of all trending videos and around 35% of recently uploaded videos are shorter than one minute. In general, the percentage of videos decreases as the duration increases; meanwhile, only about 6% of videos of both sets are longer than 10 minutes. The results show that there is not a significant difference in duration distribution of trending and recently uploaded videos. It is useful for the websites such as channelFactory, to know that they do not need to worry about a video's duration as a factor of making it popular.

### 5.1.2 Number of views

Distribution of the number of videos with more than certain number of views is shown in Fig. 6(b). We also recorded the same distribution for both trending and non-trending videos eight month later (mid March 2013) and it is also shown in Fig. 6(b). Based on the recorded results of July and March, we observe that about 90 percent of trending videos received at least 10,000 views while all of our recently uploaded videos have achieved less than 10,000 views. We also observe that 15% of our collected trending videos have achieved more than 1,000,000 views only in couple of days after they were uploaded. It is shown that the recorded distributions in March 2013 are almost the same as those in July 2012 and during the next eight months, none of the non-trending videos gained any major attention. Therefore, the non-trending videos' views get saturated very fast and to gain a better insight into the viewership of videos, Fig. 7(a) shows the normalized mean of cumulative distribution of both datasets.We observe that recently uploaded videos reach more than 90% of their views within the first two days; meanwhile the number of views of trending videos continues to grow over a much longer duration of time. Therefore, as we said before, the non-trending videos' views saturates in the first couple of days after they got uploaded and then, they will not expect any major attention except for sporadic increase in their viewership. These results demonstrate the potential of how trending videos can become phenomenally popular in comparison with recently uploaded ones that were not labeled trending. For example, one of the trending videos in our dataset reached more than 120 million views in couple of weeks.

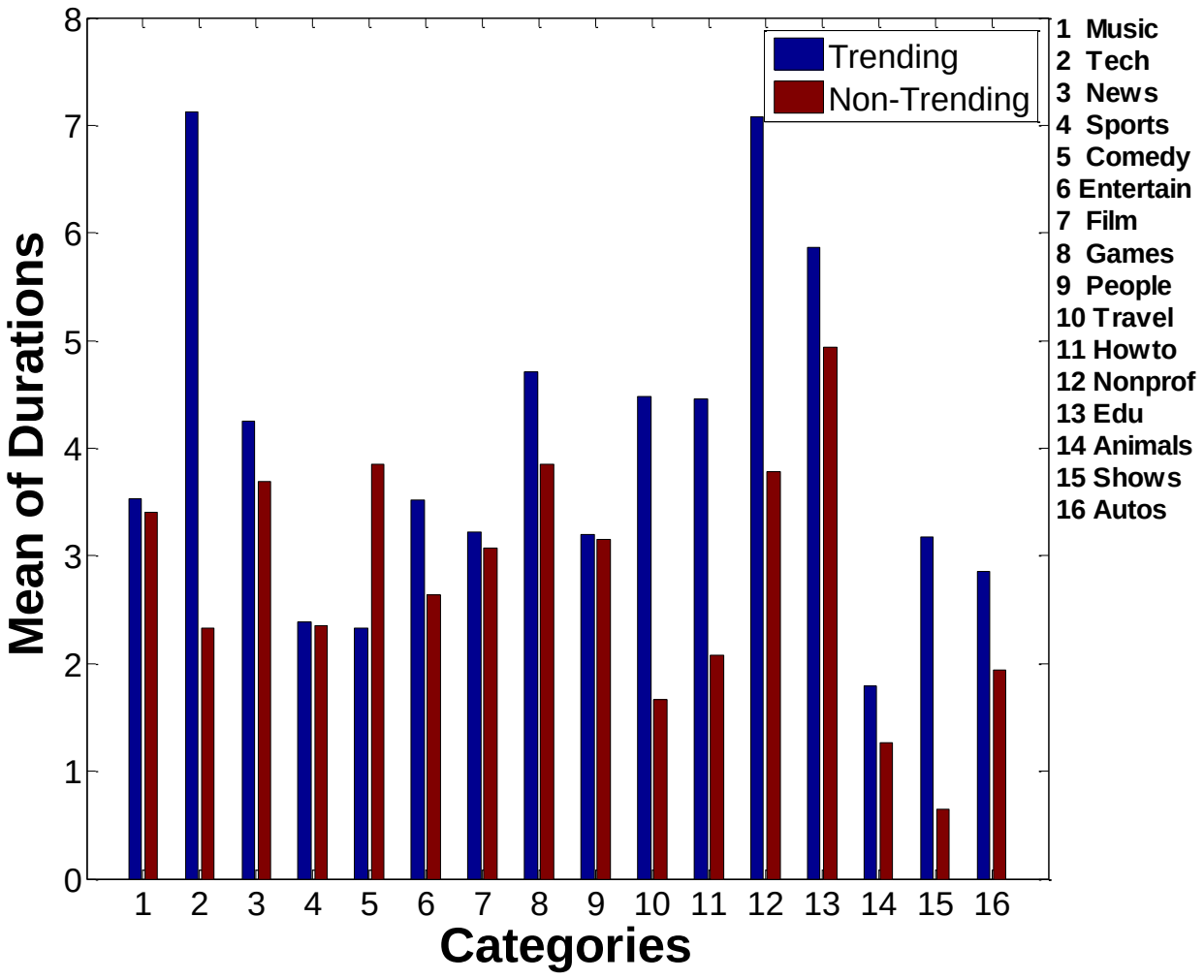

**Figure 5. Mean of durations (in minutes) of trending and recently-uploaded videos under different categories**

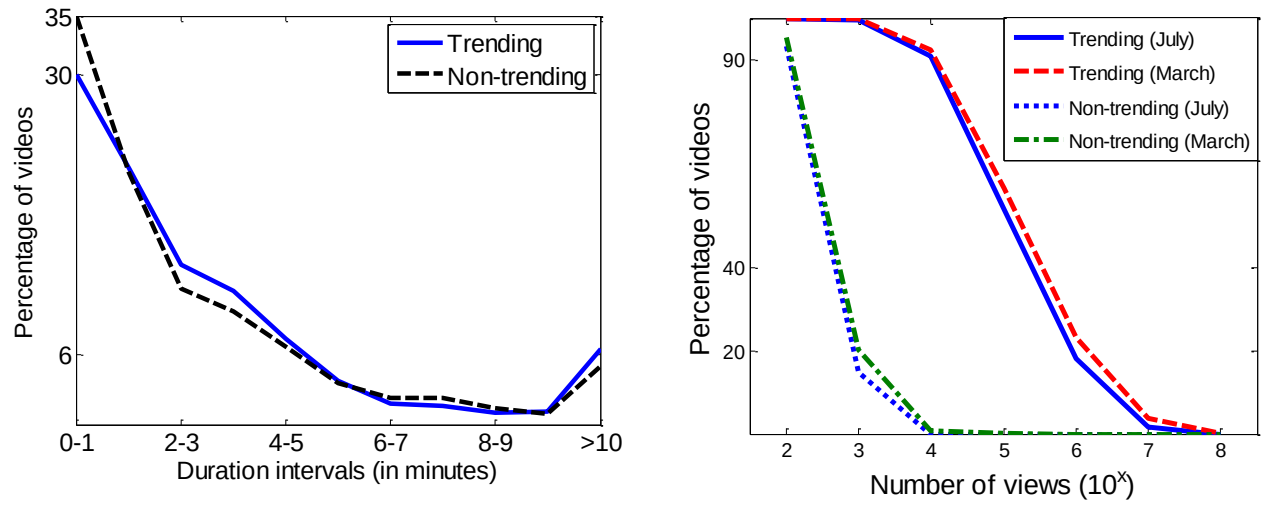

(a) (b)

**Figure 6. (a) Percentage of videos in different duration intervals. (b) Percentage of videos with more than certain number of views.**

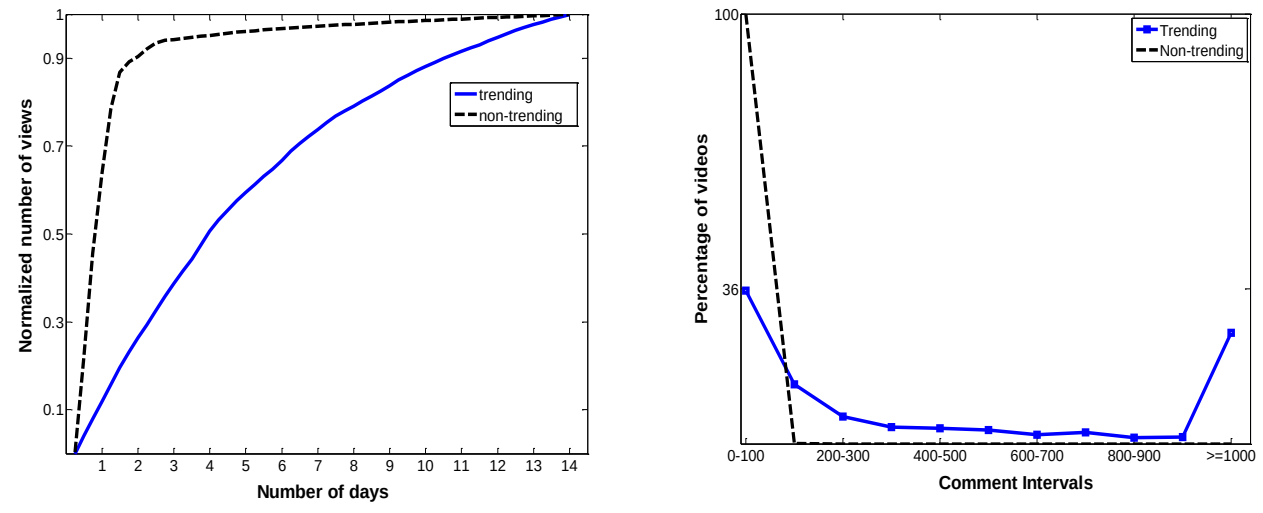

(a) (b)

**Figure 7. (a) Percentage of aggregate number of views on each day after their publish date (b) Percentage of videos for different comments interval.**

### *5.1.3 Comments*

Comments provide real insights into audience reactions to important issues or particular videos. Current research focuses on the analysis of text information of comments and tries to reveal its impact on videos' viewership [16]. We monitored the number of comments for all trending and recently uploaded videos as well. As shown in Fig. 7(b), every recently uploaded video in our dataset has less than 100 comments. For trending videos, there are two distinct regions for the distribution (arguably a bimodal distribution). About one half of the videos received less than few hundred comments per video; the other part of the distribution shows that about 30% of trending videos have more than 1,000 comments. This bimodal behavior may be attributed to the level of popularity of trending videos; some become popular and some do not increase in popularity; whereas the vast majority of non-trending videos do not achieve high levels of popularity.

## 5.2 Comparative Analysis of User Profiles

Unlike traditional statistics such as views and comments of trending videos, in this part, we attempt to gain an insight into other factors that may influence a video to get labeled as trending, such as the uploader's profile. As the name indicates, *uploaders* are those users who make their video content available to other users by uploading that content. Here, we present and evaluate publicly available statistics of users' profiles, which contain information that uploaders themselves make available to other users.

**Table 2. (a) Gender of uploaders for Trending and Non-Trending videos. (b) Percentage of users with maximum allowable video-clip duration**

| Gender | M | F |
|---|---|---|
| Trend. | 86 % | 14% |
| Non-T | 74% | 26% |

(a)

| Time | 15.5 | 720 |
|---|---|---|
| Trend. | 47% | 53% |
| Non-T | 52% | 48% |

(b)

### *5.2.1 Gender*

In our dataset, we retrieve the uploaders' gender. From Table 2(a), we observe that 86% of the uploaders are male for trending videos and 74% for non-trending. Meanwhile, it might be interesting to note that the percentage of female uploaders for trending videos is about one-half of the female percentage of non-trending video uploaders.

### *5.2.2 Maximum allowed upload duration*

The maximum length for an uploaded video is limited by YouTube. By default, the upload duration limit is 15.5 minutes for all users. Meanwhile, uploaders, who are in good standing and without any copyright or community guidelines' violations, are able to upload longer length (up to 720 minutes) videos [17]. As shown in Table 2(b), for our collected trending videos, 53% of the uploaders can upload videos with duration up to 720 minutes, which is higher than the percentage of the non-trending videos' uploaders. These numbers show that only slightly higher than half of the trending video uploaders are certified by YouTube. Based on these statistics, one may argue that the standing of an uploader with YouTube does not seem to play a major role in the selection of its video of being trending.

### *5.2.3 Subscriber count*

A *channel* on YouTube is basically the homepage of users who have a YouTube account. A channel shows the account information, the public videos that users have uploaded and any users' information that users have entered. Any uploader needs to have an account to upload a video on YouTube. Subscriber count indicates the number of YouTube users who have subscribed to a specific user's channel. Arguably, this count is an approximation for the size of the audience of a channel. The percentage of uploaders, who have more than a certain number of subscribers, is shown in Fig. 8(a). We observe that 84 percent of uploaders of our collected trending videos have more than 100 subscribers and about 6 percent of them have more than one million subscribers for their channels; while about 16 percent of uploaders of recently uploaded non-trending videos have more than 100 subscribers and about 99 percent of them have less than only 10,000 subscribers for their channel. Regardless of the content and the characteristics of a video, uploading it through a popular channel will expose the video to a wider audience. The number of subscribers for a channel is a good indication of channel's popularity. Hence, having more subscribers for your channel by e.g. linking your channel to other popular channels or sharing your videos in other social networks, you are increasing the chance of your future videos to be watched as well as the chance of becoming trending.

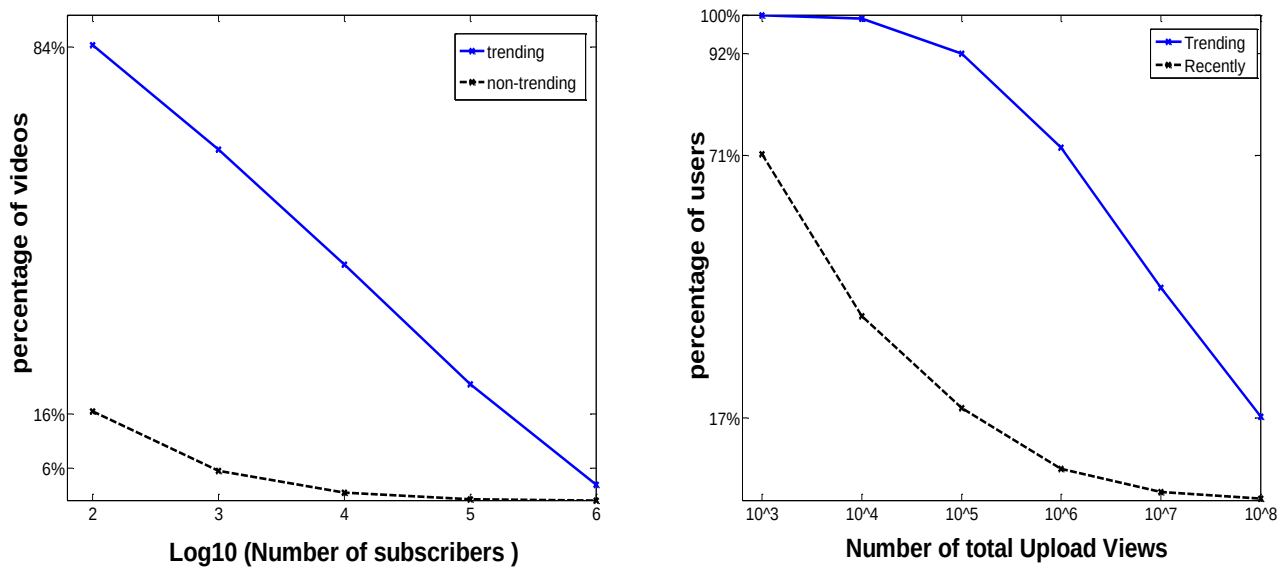

(a) (b)

**Figure 8. (a) Percentage of users with more than certain number of subscribers (b) Percentage of users with more than certain number of total uploads views**

### 5.2.4 *Total upload views*

The total upload views feed indicates the total number of views for all uploaded videos of the channel of a particular uploader. As shown in Fig. 8(b), about 99 percent of the uploaders of our collected trending videos share videos having more than 10,000 total number of views and about 17 percent of them have more than one hundred million total number of views. Only about 71 percent of uploaders of recently uploaded videos of our dataset have more than 1000 total number of views and only about one percent of them have more than ten million views. It is clear that channels associated with the trending videos of our dataset have significantly larger total upload views in comparison with the channels of non-trending videos.

### 5.2.5 *Discussion*

In summary, one can observe distinct statistical attributes that differentiate between trending and no-trending videos in our dataset and in the profiles of their uploaders. The popularity of trending videos clearly reaches significantly higher levels, and it continues to grow over longer time. Both the number of subscribers and aggregate number of viewers for the channels of trending videos are also clearly higher than their non-trending counterpart. Therefore this result is useful for those who want to expose the videos to a wider audience and try to make them popular; and not only the uploaders should worry about the content, they also need to take care of where they are going to upload the videos.

## 6. DIRECTIONAL-RELATIONSHIP ANALYSIS

As mentioned earlier, our dataset provides us with an opportunity to study and analyze the inter-relationships among trending videos' viewership through their recordings (time series). A key ultimate objective under this part of our effort is to answer the following question: Can we gain some insight about the pattern of viewership among different categories (e.g., sports, news, entertainment, etc.) of trending videos by using the time-series available to us? Addressing this question can naturally affect the strategies for marketing, target advertising, recommendation systems and search engines [2]. For example, if we know the viewership pattern of different categories and know that people tend to end up watching some categories more than others, advertisers can focus more on these categories or better marketing strategies can be found. Hypothetically speaking, answering this question can be achieved by monitoring the clicking pattern of a large number of YouTube viewers, and by tracking their transition from one video to another. This type of monitoring and tracking of viewers is arguably impossible to conduct in an automated manner due to technical, logistical, and privacy issues.

Hence, and within the context of the trending-video viewership time-series available to us, the question becomes: can we use these time-series to infer some information about the inter-relationships among different trending videos? It is important to note that one can use traditional correlation analysis to identify the (undirected) relationship between any pair of time-series. However, all we can infer from such analysis is that two time-series are correlated or not; which may imply that one can gain some information about the viewership process of one trending video by monitoring the viewership statistics of another trending video. However, this type of simplistic analysis lacks the directionality needed for addressing the aforementioned objective of inferring viewership pattern among different videos. In particular, and at the individual video-clip level, our question becomes: can we infer if viewership of a given trending video could be followed by viewership of another trending video? And we need to infer that only from the viewership time-series available to us.

Consequently, we aim at deriving *directional-relationships* among trending-video time-series rather than measuring simple (undirected) correlation among them. To measure the proposed directional-relationships among viewership time-series we resort to Granger Causality [10][11], which has been used successfully and extensively for a broad range of signal processing applications, finance, neuroscience, and political relationships [18]-[21]. Before describing Granger Causality (GC) and how it can be used in our analysis, it is important to highlight the following. Despite the term "causality" that is used by this GC measure, we caution the reader that we are not claiming that we are actually inferring from such measure that viewing a certain video *causes* viewership of another video. Although this causal effect might take place due to a variety of factors; however, we can't be certain that such causal effect is actually happening. On the other hand, Granger causality is well known to provide the directionality information (that we need) in addition to the strength of relationship between any pairs of time-series. Hence, Granger causality, and as we explain further below, provides a powerful tool for measuring the directional-relationship that we seek under this part of our effort. (For the remainder of this section, we will use the terms "directional-relationship" and "directionality" interchangeably.)

Our directionality analysis consists of three stages. First, we begin by measuring the pair-wise directional-relationship values among all pairs of viewership time-series in our original dataset of 4,000 trending videos. This provides us with the necessary tool for the second stage of our analysis: detecting if a *significant directional-relationship* (to be defined later) does exist between any given pair of trending-video time-series. Once we establish the existence or absence of a significant directional-relationship among each pair of trending video, then we can look into the categories that these videos belong to. Hence, the third and final stage is to provide directionality analysis among the different categories of trending videos. Before proceeding with our description of the GC measure and how we employed it throughout the three stages of our analysis, we provide a general perspective of directional-relationship and the implication of detecting such relationship. We believe that this perspective is crucial for assessing the viability and limitations of our directional-relationship analysis.

In general, measuring a directional-relationship between two processes $X$ and $Y$, results in two distinct values. In our context, the two processes $X$ and $Y$ represent two time-series capturing the viewership statistics of two trending videos over time. (Here, we use $X$ and $Y$ to refer to two different videos or their corresponding time-series.) Let $G_{X\to Y}$ and $G_{Y\to X}$ be the two directional-relationship values that we measure between the two processes $X$ and $Y$ using the GC measure. In general, $G_{X\to Y} \neq G_{Y\to X}$. For now, let's refer to $G_{X\to Y}$ as the directional-relationship from process $X$ to process $Y$. (Similarly, $G_{Y\to X}$ is the directional-relationship from process $Y$ to process $X$.) Before describing how to measure these two quantities, $G_{X\to Y}$ and $G_{Y\to X}$, it is important to provide a context for their meanings. More specifically, it is important to recognize the implication of having large or small values for either or both of these measures. To that end, we briefly outline the implications (or *possible* implications) of having a high value for the proposed directional measures:

1. *Predictability*: As we will see, and by definition, a high-value for $G_{X\to Y}$ implies that we can improve our prediction of future samples of $Y$ by observing past samples of $X$. However, this (high-value for $G_{X\to Y}$) does not necessarily imply that we can improve our prediction of future samples of $X$ by observing past samples of $Y$. The latter would be true if we have a high value for $G_{Y\to X}$. This highlights the importance of the directionality and the potential for asymmetry of such measures. The first stage of our directionality analysis provides a measure of the level of predictability among all trending-video time-series in our dataset.

2. *Causality*: A *plausible* implication of a high-value for $G_{X\to Y}$ in the context of our analysis of trending-video time-series is the following. Viewership of trending video $X$ has some influence (causal effect) on the viewership process for $Y$. (Please note our emphasis on the word plausible.) However, we need to develop some confidence that such causality effect might actually exist. This can be achieved by conducting significance testing on the raw $G_{X\to Y}$ value; and based on the result of this test, we can then say, with some confidence, that there is a possibility for a causal effect from $X$ to $Y$. Thus, it is important to note that predictability does not necessarily imply causality. The second stage of our directionality analysis is to assess if we can conclude, with some confidence, that there is causality between two viewership time-series.

3. *Transitionality (Group-Directionality)*: Once we establish with some confidence the presence of a causal relationship among a collection of time-series, we can look into grouping the relevant videos onto their categories. Let video $X$ belongs to category $\chi$ and let video $Y$ belongs to category $\Upsilon$. Hence, we have $X \in \chi$ and $Y \in \Upsilon$. Detecting an isolated causal effect from $X$ to $Y$ does not necessarily imply that there is a strong directional relationship between category $\chi$ and category $\Upsilon$. To measure the directional relationship among different groups (or video categories in our case) we look into the total causal effect from time-series in category $\chi$ toward time-series in category $\Upsilon$. In other words, we add all of the significant directionality scores (that pass the significance testing from the second stage of our analysis) to measure the overall directional-relationship $\mathcal{T}_{\chi\to\Upsilon}$ from category $\chi$ toward category $\Upsilon$. We refer to this group-level directionality measure $\mathcal{T}_{\chi\to\Upsilon}$ as *transitionality*. The proposed transitionality measure $\mathcal{T}_{\chi\to\Upsilon}$ can have the following interpretation in the context of our problem. Viewers of trending videos belonging to category $\chi$ might have a tendency to watch one or more trending videos that belong to group $\Upsilon$ (after watching one or more trending videos in group $\chi$). This interpretation implies some form of a transitional behavior: viewership of $\chi$ tends to transit to viewership of $\Upsilon$. We refer to this plausible transitional behavior as *transitionality*. In essence, transitional-tendency is reminiscent, in some respects, of a Makov chain model of a random process. As we will see from our analysis results, we have high directional-relationship values toward a popular category, say $\Upsilon$, from virtually all other categories $\chi_1$, $\chi_2$,... etc. We believe that such scenario can be explained with a group-level transitional tendency interpretation. In other words, it seems quite plausible that regardless which video category a viewer may watch, then that viewer tends to watch a popular category (e.g., music or entertainment) afterword.

In the remainder of this section, we (a) describe the GC measure and how we employ it in our directional-relationship analysis and (b) present the results of our analyses at the individual trending-video time-series level and at the video category level.

### 6.1 Granger Causality

Granger causality (GC) is widely used to describe the directional-relationship (or causality) between two time series. Traditional measures, such as correlation, coherence, and mutual information, are only able to quantify the strength of the relationship between two random processes; Granger causality can capture both the strength and direction of information flow between two random processes [10][11]. Under the context of GC, a stochastic process $X$ is considered causing another process $Y$ if the prediction of $Y$ at the current time point, $Y_t$, is improved when taking into account the past samples of $X$. Granger causality is commonly implemented within a linear prediction framework using a bivariate autoregressive (AR) model [12]. In this framework, signals are fitted by both univariate and bivariate autoregressive models; the improvement of the prediction for $Y_t$ is assessed by the difference of the variance of the prediction error between these two models. When each process is fitted in a univariate signal model, the prediction of the current sample of $Y_t$ only depends on the past samples of itself as the equations shown below,

$$X_t = \sum_{i=1}^{p1} a_{xi} X_{t-i} + n_{xt} \tag{1}$$

$$Y_t = \sum_{i=1}^{p2} a_{yi} Y_{t-i} + n_{yt} \tag{2}$$

Where $p1$ and $p2$ are the order of the random processes $X$ and $Y$ respectively, $a_{xi}$ and $a_{yi}$ are the autoregressive coefficients, and $n_{xt}$ and $n_{yt}$ are the noise at time point $t$. When each signal is fitted in a bivariate AR model, i.e.,

$$X_t = \sum_{i=1}^{p1} a_{xi} X_{t-i} + \sum_{i=1}^{p3} b_{xi} Y_{t-i} + n_{xt} \tag{3}$$

$$Y_t = \sum_{i=1}^{p2} a_{yi} X_{t-i} + \sum_{i=1}^{p4} b_{yi} Y_{t-i} + n_{yt} \tag{4}$$

The prediction of each signal depends on the past sample of both signals.

Granger employs variance to evaluate the improvement of the prediction error of $Y_t$ and the Granger causality from $X$ to $Y$ can be quantified as:

$$G_{X\to Y} = \ln\left(\frac{var(Y_t|Y^{t-1})}{var(Y_t|X^{t-1}Y^{t-1})}\right) \tag{5}$$

Here, $Y^{t-1}$ ($X^{t-1}$) is the past $t-1$ samples of $Y_t$ ($X_t$). If $var(Y_t|Y^{t-1}) > var(Y_t|X^{t-1}Y^{t-1})$, $X$ has a causal effect on $Y$; if the past of $X$ does not improve the prediction of $Y_t$, $G_{X\to Y}$ is close to or less than 0. In this paper, the MATLAB toolbox developed by Seth is used to compute the Granger causality value [13]. Granger causality is normalized to the $[0, 1]$ range for comparison purposes.

### 6.2 Significance Testing

The obtained GC value does not mean there is significant causal effect between random processes $X$ and $Y$. We employed bootstrapping method to test the significance of the obtained GC value. In order to test the null hypothesis of noncausality, for each time series, we randomized the order of all time points of the time series $X$ 100 times to generate new observations $X_m^*$, $m = 1, \dots, 100$; in this way, the causal structure between $X$ and $Y$ is destroyed [14]. We compute the GC value for each pair of random processes ($X_m^*$ and $Y$). A threshold $G_{th}(\alpha)$ is obtained at a significance level of $\alpha$ (e.g., $\alpha = 0.05$), such that $(1-\alpha)\%$ (i.e., 95%) of the GC values for randomized pairs of data ($G_{X_m^*\to Y}$) are less than this threshold, $G_{X_m^*\to Y} < G_{th}(\alpha)$. If the GC value $G_{X\to Y}$ of the original pair of data ($X$ and $Y$) is larger than this threshold, i.e. if $G_{X\to Y} > G_{th}(\alpha)$, then this indicates that there is a *significant directional-relationship* from $X$ to $Y$. Hence, we distinguish between two types of values: (a) A raw directional-relationship value that is based on measuring $G_{X\to Y}$; we refer to this value by the *raw* GC measure, and it provides a direct measure for predictability. (b) A significant directionality value that is based on performing the significance testing on the raw GC value; we refer to this second type by the *significant* GC measure $SG_{X\to Y}(\alpha)$, which provides a measure for causality from $X$ to $Y$.

### 6.3 Data Processing

In order to apply Granger causality to our dataset, we had to preprocess the data as follows. First, since the list of trending videos is continuously changing over time, the time duration over which we were able to monitor each trending video statistics is different for each video clip. Therefore, we had to choose trending videos that we were able to monitor over the same duration of time to make sure that all time series having the same start and end points. This alignment is crucial for directional-relationship analysis. As mentioned earlier, we used the time-series of our first dataset that consists of about 4,000 trending videos (3922 to be exact). More importantly, we had to ensure that we select a time duration when these videos reach some level of maturity in terms of their popularity. Consequently, and over the total of eight weeks that we monitored these videos, we selected the time-segments (of the 4,000 time-series) covering the last week for our directional-relationship analysis. During that last week: (a) all 4,000 videos have time-series covering that week; and (b) all 4,000 videos have reached a significant level of their popularity. It is important to note that our directionality analysis in this paper does not cover (actually tries to avoid) any longitudinal effect over time. Such effect could be the subject of a future paper. Therefore, we consider our directionality study as a "snapshot in time" analysis.

Finally, even over a one week-period, the time series exhibited a non-stationary behavior (as can be observed in Fig. 9(a)), which may affect the estimation of Granger causality. In order to reduce the effect of non-stationarity, we subtract the best-fitting line from each time series and remove the temporal mean from each observation of the time series [13]. The time series with zero-mean were used in our model fitting and Granger causality computation.

## 6.4 Directionality Using the Granger Measure

Once we preprocess the data, we are able to evaluate the GC measures among all time-series pairs in our dataset. A crucial observation that we made early on during our effort is that meaningful directionality analysis can only be achieved by dividing our dataset into subsets based on the level of viewership (popularity) of the videos. For example, some trending videos may only receive an aggregate of thousands of views after several weeks of being uploaded, while others receive millions of views. In that case, one would expect very small directional relationship between a video that has more than one million views and another video with less than one thousand views. Furthermore, considering all videos without any regard to their popularity leads to highly non-stationary processes as mentioned above. The mean and variance of the hourly viewership of all videos over the period of our directionality analysis are shown in Fig. 9. We observe that the variance of the hourly viewership is very large, which may lead to inaccurate directionality and causality analysis results. Hence, we divided our dataset of 4,000 videos into different subsets based on the popularity of these videos.

The histogram of the log value of aggregated viewership of all 3,922 videos at the end of two months recordings is shown in Fig. 1(a). We observe that the viewership covers a wide range. To investigate the relationship between causality and popularity, we computed the pairwise raw GC value $G_{X \to Y}$ among all videos in our dataset; and grouped these videos into five different popularity levels based on their final aggregated viewership according to Fig. 1, i.e. the aggregated viewership is in the ranges of $10^3 \sim 10^4$, $10^4 \sim 10^5$, $10^5 \sim 10^6$, $10^6 \sim 10^7$, and $10^7 \sim 10^8$. We do not consider the effect of video in the range of $10^2 \sim 10^3$, since there is only one video in this range. The value we assigned for the directional-relationship between two popularity levels is the averaged raw GC values between all pair-wise time-series of videos in these two different popular levels. The results are shown in Fig. 10. The rows in the figure represent the source and the columns represent the destination popularity levels.

In general, we observe that the GC directional-relationship values from popular videos to less popular videos are larger than the values from less popular videos to more popular videos. For example, videos with viewership in the range of $10^7 \sim 10^8$ have strong impact on videos with less viewership; while the opposite is not true. Also, one can observe that directional GC values are smaller among groups with different popularity levels; whereas the GC values tend to be larger among groups with similar popularity values. Furthermore, the GC values among trending videos that do not become popular was also lacking. Hence, and overall, we found that noticeable causal effect can be observed among trending videos that become popular; in particular those videos having more than 1,000,000 views. Consequently, for the remainder of this section, we mainly focus on presenting our directionality analysis on this group of videos, which consists of 321 time-series. We also present results for the set of videos having more than 5,000,000 aggregate views. The primary motivation for presenting results regarding the latter set with five million-plus views is that they consist of 62 videos; this enables us to illustrate some of our results visually (at the individual video clip level) using causality matrices as shown below. Afterword, we revert back to presenting our directionality analysis for the one-million-plus set of 321 videos. It is important to note that the analysis presented below can be applied to any group of videos in our dataset or other datasets. Our focus on the two sets with one-million-plus and five-million-plus views are mainly intended as a vehicle for presenting the remainder of our directionality analyses.

## 6.5 Causality Analysis

As we discussed above, we focus on two subsets of trending videos. The first subset consists of 62 trending videos having more than five million views (mainly for illustrative purposes). We computed the GC value between every two videos within this group. The obtained raw GC values are shown in Fig. 11 (a). It is important to note the following. First, the GC values exhibit a clear asymmetric patter. This emphasizes the importance of conducting directionality analysis among the time-series of these videos. Second, the GC values are rather high (mostly around or higher than 0.5). By definition, this implies a good level of predictability among many pairs of viewership time-series, especially among those with very high GC values. However, these raw GC values do not necessarily translate into causal effects among the corresponding time series.

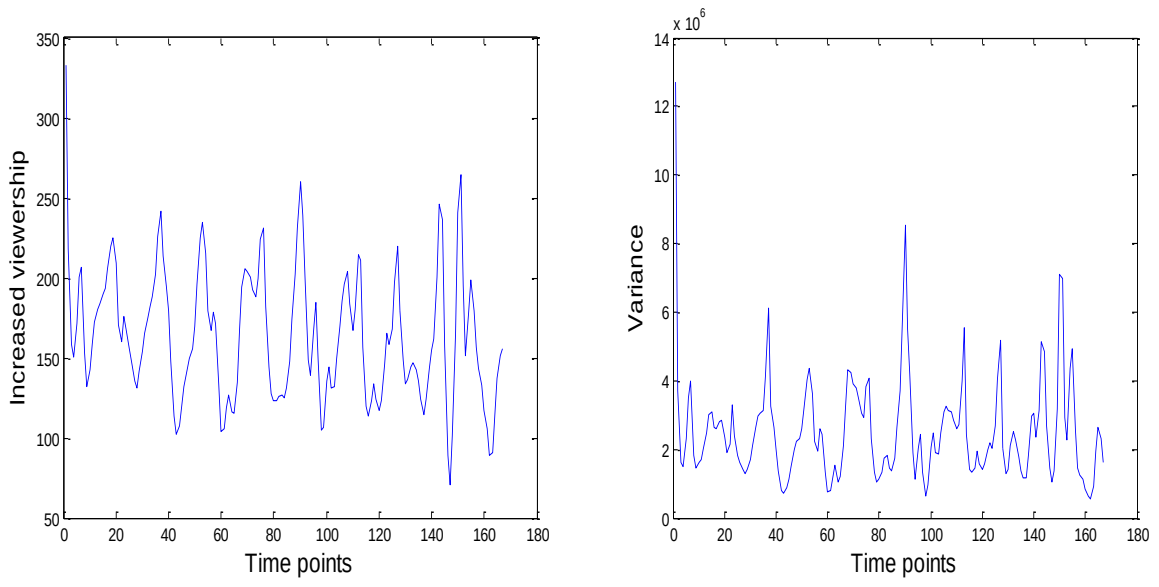

(a) (b)

**Figure 9. (a) Average hourly views of trending videos in our database. (b) Variance of hourly views for all 3922 videos in one hour.**

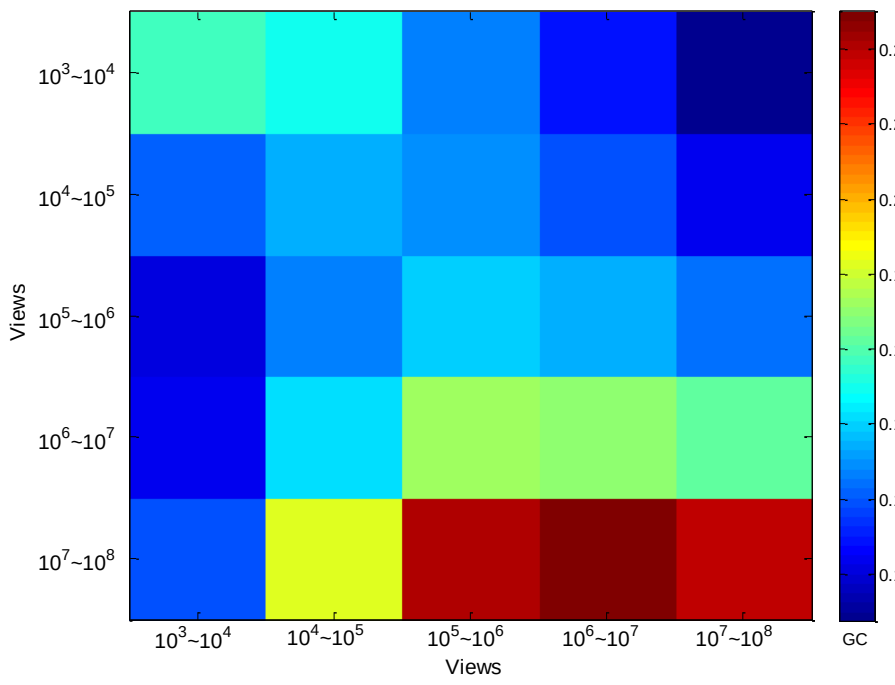

**Figure 10. The directional-relationships among different popularity levels using average GC values.**

In order to test the significance of the pair-wise GC values in terms of causality, we obtain a threshold at 5% significance level for each pair-wise entry value in the GC matrix of Fig. 11(a). We used the bootstrapping method (described earlier) for each GC value. The video pairs that have significant GC values are shown with their corresponding strength in Fig. 11 (b). The results show that 192 of all 62*62 pair-wise GC values are significant. Hence, for these directional-relationships, we are 95% confident that there is some level of causality among each pair that passed the significance test. We notice that virtually all significant GC values that survived have very high raw GC values to begin with; these significant values are mostly around 0.9 while their minimum level is above 0.7. Moving forward, we can discuss the causality among these time series that exhibit significant GC values.

Based on the pair-wise causal information, we are able to show (a) the causal effect (or influence) of the viewership of each particular video ***on*** the viewership of other videos; and (b) the level of influence that each video experiences ***from*** the causal effect being exerted on it by other videos. These two cases are illustrated in Fig. 12 (a) and (b), respectively. The percentage of videos that influence more than $m$ percentage ($x$ axis) of all 62 trending videos is shown in Fig. 12 (a). We observe that around 90% of all videos influence the viewership of at least one other video. The most influential video exhibit causal effect on more than 16% of all videos; meanwhile, the majority of videos influence less than 10% of all videos. We also uncovered how much the viewership of one particular video is influenced by the viewership of the other videos (case (b) above). From Fig. 12 (b), we observe that more than 70% of videos are influenced by at least one video. The most influenced video is impacted by the viewership of more than 20% of all videos.

For each specific video, the percentage of videos it affects (blue dots) and the percentage of videos that affects this particular video (red triangles) are shown in Fig. 12 (c). (We sorted the values of the blue dots for illustrative purposes only.) We observe many 'active' (only influence others) and 'passive' (only influenced by others) videos; other videos seem to fall between these two categories. There is only one isolated video (video 4), which neither influences nor gets influenced by other videos. It is probably because the viewership of this video is near saturation or does not change much in our chosen time period, and hence, has little influence on the viewership of other videos. Another possibility is that it may have directional relationships with videos that have less than 5 million views.

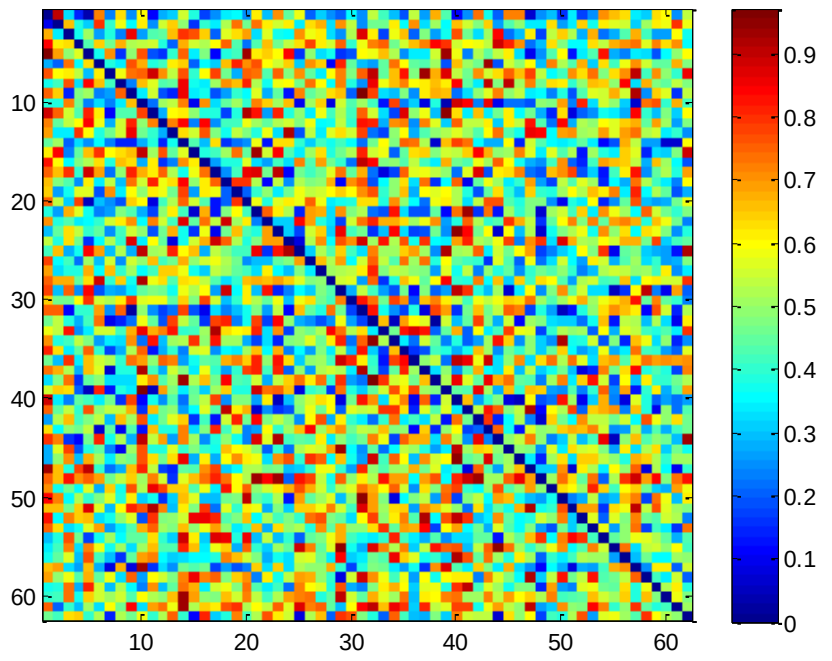

**(a) Pair-wise Granger causality (raw values)**

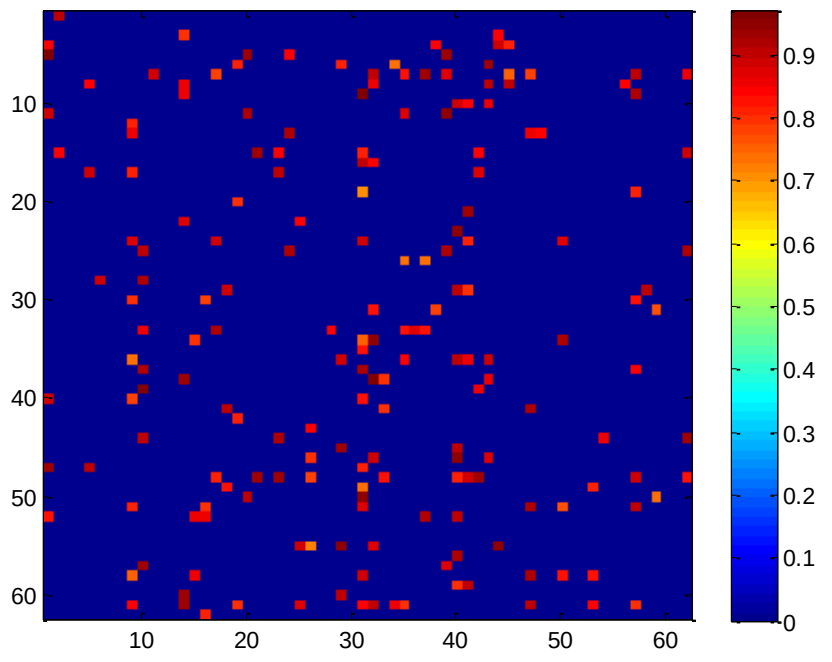

**(b) Significant GC pairs (after significance testing)**

**Figure 11. Causality analysis of videos with more than 5 million views.**

As we mentioned above, to compare the above results with another group of videos with a different level of popularity, we did the same directionality/causality analysis for trending videos having more than one million views (321 videos). The results are shown in Fig. 13. All videos affect at least one video; the most influential videos only influence 16% of all other videos. The percentage (influence pattern) is nearly the same with videos having more than 5 million views. It is also interesting to see that 2% of our selected videos are not influenced by the other videos; the viewership of the most influenced video is impacted by 16% of all other videos. The results, shown in Figs. 12 and 13 for different datasets at the same time interval, indicate that the viewership of a particular trending video is influencing or influenced by at least one other trending videos (except the isolated video – number 4 – in Fig. 12 (c)).

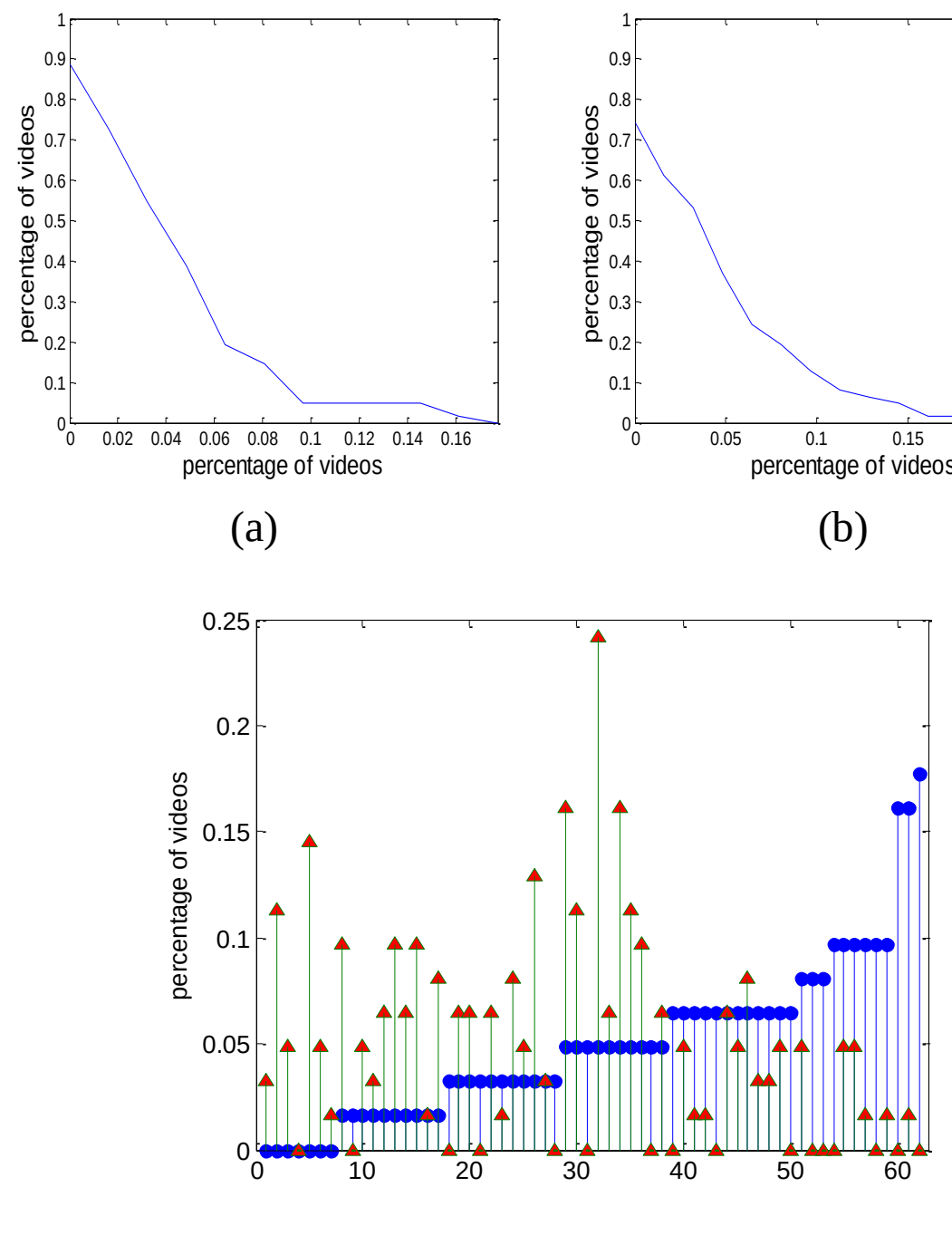

**Figure 12. 62 trending with 5-million-plus views. (a) The percentage of videos which influence more than m (x axis) percentage of total videos. (b) The percentage of videos which is influenced by more than m (x axis) percentage of total videos. (c) For each specific video, the percentage of videos it affects (blue dot) and the percentage of videos it is influenced by (red triangle).**

## 6.6 Transitionality Analysis

The causality analysis reveals both the strength and direction of how the viewership of two videos affects each other. If the Granger causality value from video $i$ to $j$ is significant, where $i, j = 1, \dots, N$ with $N$ being the number of videos, it implies that people may watch video $j$ through video $i$.

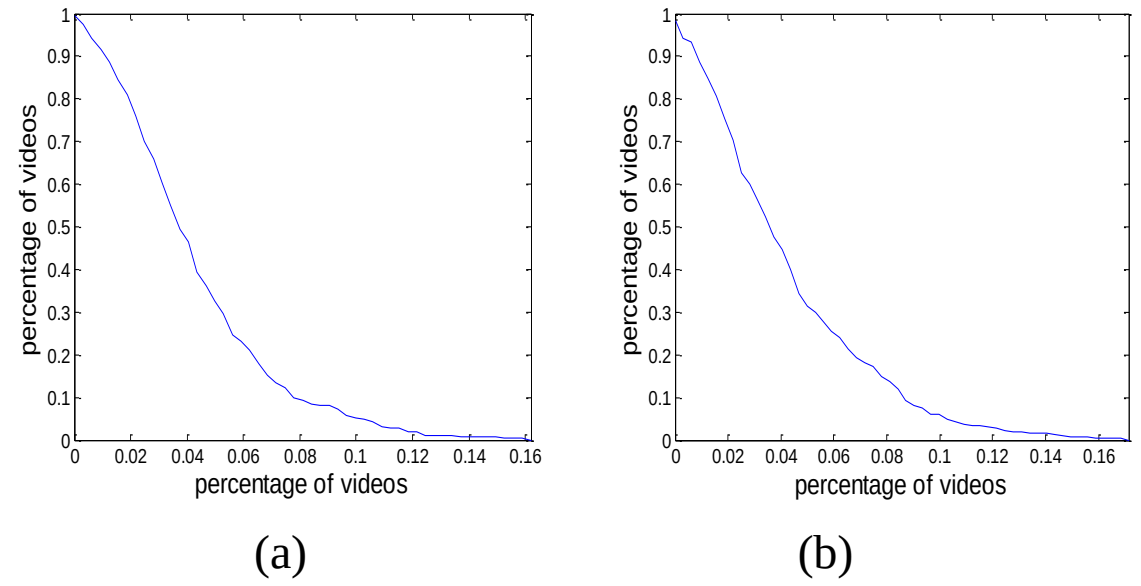

**Figure 13. 321 trending videos with more the one million views. (a) The percentage of videos which influence more than m (x axis) percentage of total videos. (b) The percentage of videos which is influenced by more than m (x axis) percentage of total videos.**

However, this implication is rather video-specific; it would be more interesting to reveal the underlying viewership pattern among different trending videos' categories. We refer to a directional-relationship among different groups (categories in our case) as transitionality (as discussed earlier).

In our transitionality analysis, we focus on videos having more than one million aggregated views (as opposed to the set with more than five million views). A directional-relationship value from category $\chi$ to category $\Upsilon$ is assigned a transitionality score $\mathcal{T}_{\chi\rightarrow\Upsilon}$ as follows:

$$\mathcal{T}_{\chi\rightarrow\Upsilon} = \sum_{Y\in\Upsilon}\sum_{X\in\chi} SG_{X\rightarrow Y}$$

Here, $SG_{X\rightarrow Y}$ represents the significant GC values derived from the Granger causality analysis with significance testing as described above. Now, we can view the transitionality among the different categories using a weighted graph as shown in Fig. 14. The weight of each link represents the transitionality score computed by the above formula. In this example, the weights of these links can be as high as 240, indicating that there could be about 240 significant causality links from one category to another. However, many inter-category links have transitionality scores lower than 20, which are not shown in the figure.

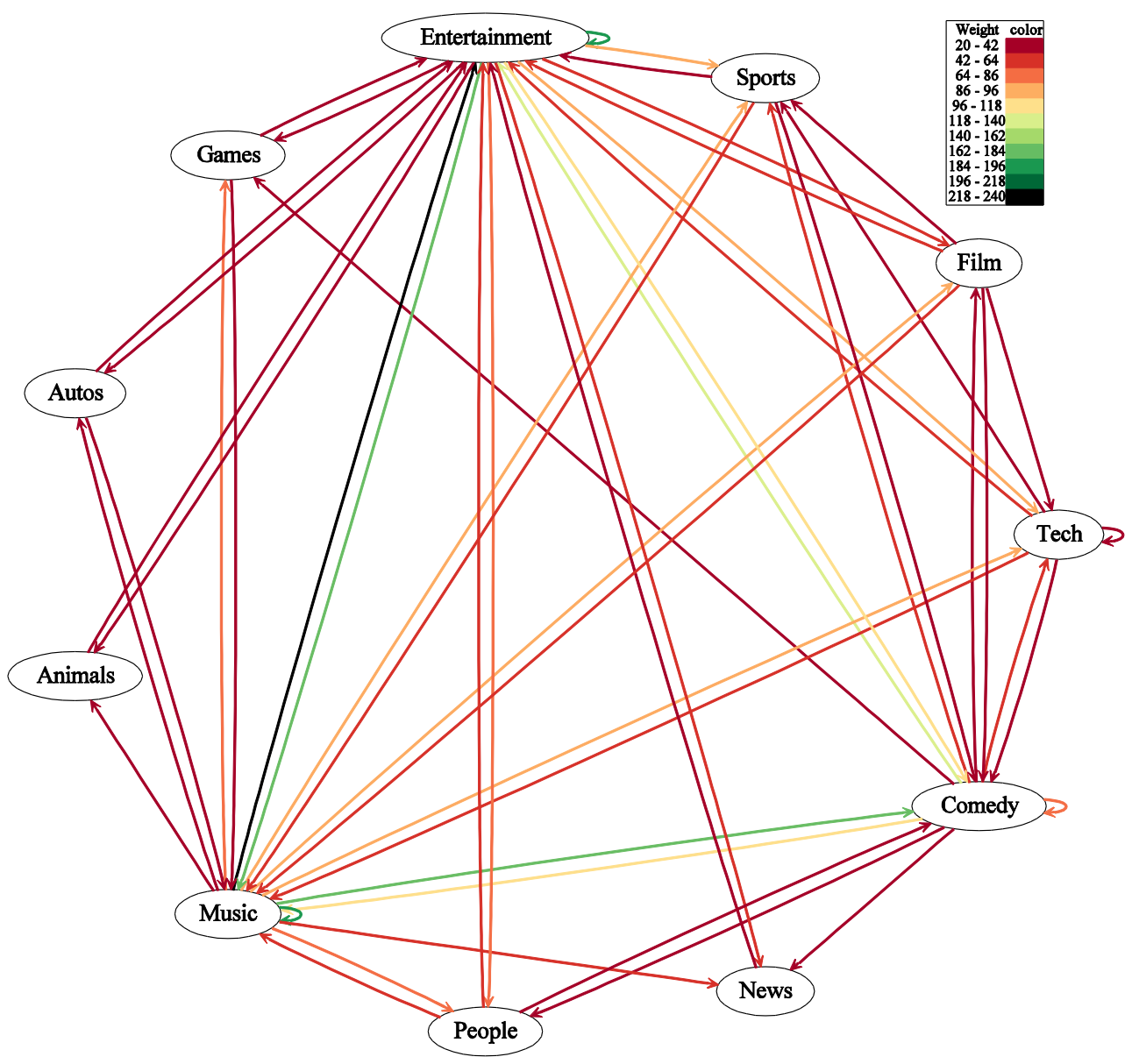

**Figure 14: Clicking pattern of users that watch different trending videos' category. Videos having more than one million views are considered. This figure, being simplified for an easier presentation, is excluding these five isolated categories: Howto, Nonprofit, Education, Travel and Shows.**

There are several observations can be made from this network graph. First, it is quite clear that both the entertainment and music categories have transitionality links coming into them from all other categories. This might imply that regardless which category a viewer may visit, then viewership of one or both of these categories are quite likely. Second, there is an intra-category link with a high transitionality score for each of these two popular categories; this indicates that there is also a strong tendency that someone watching a video with a popular category may tend to stay in the same category and watch another member of its videos. Third, the comedy category is close in its popularity to both entertainment and music. It is rather interesting to observe that the inter-category links among these three popular categories have relatively high transitionality scores.

Five nodes are isolated, such as Howto, Nonprofit, Education, Travel, and Shows, because their links to the other categories are weak (less than the mean value), which implying that users watch videos related to these five categories have low possibility of watching videos in other categories.

## 7. CONCLUSION

In this paper, we presented our findings for measuring, analyzing, and comparing key aspects of YouTube trending videos. Our study has been based on monitoring the viewership and related statistics of more than 8,000 YouTube videos over an aggregate period of about three months. Since trending videos are declared as such just several hours after they are uploaded, we are able to analyze trending videos' time-series across critical and sufficiently-long durations of their lifecycle. We presented an extensive data-driven analysis on the lifecycle of trending videos. To the best of our knowledge, this work is the first study on the analysis of trending videos' lifecycles. We also presented the basic characteristics of trending videos popularity over their lifetime. In addition, we analyzed the profile of users who upload trending videos. Furthermore, we conducted a directional-relationship analysis among all pairs of trending videos' time-series that we have monitored. We employed Granger Causality (GC) with significance testing to conduct this analysis. Our GC-based directional-relationship analysis provided a deeper insight onto the viewership pattern of different categories of trending videos. Key findings of our study include the following. Trending videos and their channels have clear distinct statistical attributes when compared to other YouTube content that has not been labeled as trending. Based on the GC measure, the viewership of nearly all trending videos has some level of directional-relationship with other trending videos in our dataset. Our results also reveal a highly asymmetric directional-relationship among different categories of trending videos. Our directionality analysis also shows a clear pattern of viewership toward poplar categories, whereas some categories tend to be isolated with little evidence of transitions among them.